\documentclass{aa}
\usepackage{graphicx}
\usepackage{txfonts}
\usepackage{natbib}
\usepackage{color}

\bibpunct{(}{)}{;}{a}{}{,}

\begin{document}

\title{Are jets in symbiotic stars driven by magnetic fields?}

\author{Matthias Stute\inst{1} \and Max Camenzind\inst{1}}
\institute{Landessternwarte Heidelberg, K\"onigsstuhl, D-69117 Heidelberg, 
Germany}

\offprints{Matthias Stute, \email{M.Stute@lsw.uni-heidelberg.de}}

\date{Received 25 November 2004 / Accepted 26 January 2005}

\abstract{
We compare two scenarios to launch jets -- formation by MHD processes or 
formation by thermal pressure in the boundary layer (BL) -- with respect to 
their compatibility with observational data of jets in symbiotic stars, 
especially in the well studied jet source MWC 560. Finally, we discuss points 
of further research to be done.
}

\maketitle

\section{Introduction}

Although jets are ubiquitous phenomena in many different astrophysical 
objects as young stellar objects where they are driven by protostars, 
symbiotic stars (white dwarfs), X-ray binaries (neutron stars and stellar mass 
black holes) and active galactic nuclei (supermassive black holes), their 
formation is relatively unclear. 

The mass loss rate of the jet is found to be connected to the mass accretion 
rate of the underlying disc found in most objects \citep[e.g.][]{Liv97}. 
Therefore the necessary components seem to be well known and common to all 
objects. A more careful investigation of one specific class of objects should 
promise new insights also for the mechanisms in the other classes. From the 
observational point of view, one needs observations with a high spatial 
resolution and kinematic informations from regions as near as possible to the 
jet source. These aspects make the class of symbiotic stars ideal testbeds.

Symbiotic stars are interacting binaries with orbital periods in the range of 
years. These systems show outbursts similar to classical novae. The stellar
component is a cool red giant (RG), the hot component a white dwarf (WD) with
temperatures of 50000 -- 200000 K. Both stars show mass loss through 
supersonic winds. Wind material from the RG is captured by the WD to form an 
accretion disc. The accretion then causes thermonuclear explosions of the WD 
surface leading to an increase in luminosity followed by jet emission. Jets 
are detected in 10 out of almost 200 symbiotic stars \citep{BSK04} and this 
process was directly observed in CH Cygni \citep{TSM86}. Other famous systems 
are R Aquarii and MWC 560. While the first two objects are seen at high 
inclinations, the jet axis in MWC 560 is practically parallel to the line
of sight.

Another class of accreting WDs are cataclysmic variables (CV) which are 
{\em very close} WD binaries with a low mass main sequence star as 
companion. The mass transfer is due to Roche lobe overflow, not due to wind 
accretion. The short timescales in these systems make them the best understood 
accreting systems. Remarkably, they show no jet emission. Further related 
objects are supersoft X-ray sources, in which the temperature and pressure in 
the boundary layer (BL) of the WD are in the correct range to maintain steady 
nuclear burning on the WD surface. 

In jet formation models presented so far, the magnetic field seems to play a 
key role. The first analytical work studying magneto-centrifugal acceleration 
along magnetic field lines threading an accretion disc was done by 
\citet{BlP82}. They have shown the braking of matter in azimuthal direction 
inside the disc and their acceleration above the disc surface by the poloidal 
magnetic field components. Toroidal components of the magnetic field then 
collimate the flow. Numerous semi-analytic models extended the work of 
\citet{BlP82}, either restricted to self-similar solutions and their geometric 
limitations \citep[e.g.][]{PuN86,VlT98,VlT99,FeC04} or with non-self-similar 
solutions \citep[e.g.][]{Cam90,PeP92,BrC00}.

Another approach is to use time-dependant numerical MHD simulations to 
investigate the formation and collimation of jets. In most models, however, 
a polytropic equilibrium accretion disc was regarded as a boundary condition 
\citep[e.g.][]{KLB99,KLB04,ALK,GBW99}. The magnetic feedback on the disc 
structure is therefore not calculated self-consistently. Only in recent years 
were the first simulations including the accretion disc self-consistently in 
the calculations of jet formation presented \citep[e.g.][]{CaK02,CaK04,KMS04}. 

Due to the fact, that strong magnetic fields have been detected so far 
only in one symbiotic system \citep[Z Andromedae,][]{SoB99}, the jet formation 
by magneto-centrifugal forces exclusively seems to be insufficient. Radiative 
launching can be excluded due to too small radiation fields. A new possibility 
to accelerate plasma close to the central object was proposed involving 
SPLASHs ({\em SPatiotemporal Localized Accretion SHocks}) in 
the BL \citep{SoR03}. Locally heated bubbles expand, merge and accelerate 
plasma to velocities larger than the local escape velocity. \citet{SoL} applied
this model to disk-accreting white dwarfs to explain the absence of jets in 
CV. They have found a critical accretion rate of $\sim 10^{-6}$ 
$M_{\odot}$ yr$^{-1}$ below which no jets should be present. This scenario was 
introduced only in analytic estimates. 

In section \ref{sec_estim}, we give general estimates based on observations of 
the well studied jet in MWC 560. After that we calculate the magnetic field 
near the white dwarf required for jet formation by MHD processes in section 
\ref{sec_magn}. In section \ref{sec_obs}, we list different ways with which a 
magnetic field in MWC 560 could be detected. Finally a discussion is given.

\section{General estimates} \label{sec_estim}

As observed by \citet{SKC01} and as used to simulate the jet nozzle in 
\cite{SCS05}, the parameters of the jet in MWC 560 are 
\begin{itemize}
\item the velocity $v_{\rm jet} = 1000$ km s$^{-1}$ ,
\item the number density $n_{\rm jet} = 5 \times 10^6$ cm$^{-3}$, which is 
equal to a mass density $\rho_{\rm jet} = 8.4 \times 10^{-18}$ g cm$^{-3}$, and
\item the jet radius $R_{\rm jet} = 1$ AU.
\end{itemize}
\noindent
Using the equations  
\begin{eqnarray} 
\dot M_{\rm jet} &=& \pi \, R_{\rm jet}^2 \, m_{H} \, n_{\rm jet} \, 
v_{\rm jet} \nonumber \\
\dot P_0 &=& \dot M_{\rm jet}\,v_{\rm jet}  \nonumber \\
L_{\rm jet} &=& \frac{1}{2} \, \dot M \, v_{\rm jet}^2 = \frac{1}{2} \, \pi \,
 R_{\rm jet}^2 \, m_{H} \, n_{\rm jet} \, v_{\rm jet}^3 , \nonumber
\end{eqnarray}
this specifies 
\begin{itemize}
\item the mass outflow rate $\dot M_{\rm jet} = 9.33 \times 10^{-9}$ 
$M_{\odot}$ yr$^{-1}$ ,
\item the momentum discharge $\dot P_0 = 5.93\times 10^{25}$ g cm$^{-1}$ 
s$^{-2}$ and
\item the kinetic jet luminosity $L_{\rm jet} = 2.93\times 10^{33}$ erg 
s$^{-1}$.
\end{itemize}
\noindent
This luminosity can be provided by several different mechanisms: 
\begin{itemize}
\item by the (insufficient) luminosity of the white dwarfs 
\begin{eqnarray}
L_{\rm WD} &=& 4\,\pi\,R^2_{*}\,\sigma\,T^4 = \nonumber \\
&& 3.5 \times 10^{30} \,
\left( \frac{R_{*}}{7\times 10^8 \,{\rm cm}} \right)^2\,\left( 
\frac{T}{10^4\,{\rm K}} \right)^4 \,{\rm erg \,s}^{-1}
\end{eqnarray}
\item the accretion power of the disk (and almost the same amount coming from 
a BL) of a rotating magnetized WD
\begin{eqnarray}
L_{\rm acc} &=& \frac{G\,M_{*}\,\dot M}{R_{*}} = 1.19 \times 
10^{35} \,\left( \frac{M_{*}}{M_{\odot}} \right) \nonumber \\
&& \times\,\left( \frac{R_{*}}{7\times 10^8 \,{\rm cm}} 
\right)^{-1}\,\left( \frac{\dot M}{10^{-8}\, M_{\odot} {\rm yr}^{-1}} \right) 
\,{\rm erg \,s}^{-1}
\end{eqnarray}
\item or the magnetic luminosity \citep{Came97}
\begin{eqnarray}
L_{\rm mag} &=& \frac{1}{2}\,\Omega_{*}\,\,R^3_{*}\,B_p\,B_{\varphi} = 
4.76 \times 10^{34} \nonumber \\
&& \times\,\left( \frac{\Omega_{*}}{{\rm h}^{-1}} \right)\,
\left( \frac{R_{*}}{7\times 10^8 \,{\rm cm}} \right)^3\,
\left( \frac{B_p}{{\rm MG}} \right) \,
\left( \frac{B_{\varphi}}{{\rm MG}} \right) \,{\rm erg \,s}^{-1}.
\end{eqnarray}
\end{itemize}
\noindent
The flux in the UV band was measured as $10^{-9}$ erg cm$^{-2}$ s$^{-1}$ 
\citep{MMO91} and the UV luminosity -- which is likely to be equal to the 
accretion power of the disk and the boundary layer -- is then 
\begin{equation}
L_{\rm UV} = 1.2 \times 10^{35}\,\left( \frac{d}{{\rm kpc}} 
\right)^2 \,{\rm erg \,s}^{-1}.
\end{equation}
\noindent
With a derived distance of 2.5 kpc, this suggests an minimal accretion rate 
of $6.3 \times 10^{-8}$ $M_{\odot}$ yr$^{-1}$ and therefore an ejection 
efficiency of 14 \%. 

The minimal accretion rate is below the critical accretion rate derived by 
\citet{SoL} by a factor of 16. As these authors claim an uncertainty of a 
factor of $\sim 10$ and the remaining factor could naturally arise from 
our estimates, this should not concern.

Using the observables $\dot M$, $L_{\rm jet}$ and $\dot P_0$, one can fix in 
principle the two free parameters $\chi$ and $\Gamma$ of the model of 
\citet{SoR03}. These parameters are the fractions of the mass outflow rate 
due to SPLASHs to the accretion rate and of the initial kinetic 
energy to the final kinetic energy inside a SPLASH, respectively.
Fixing the model highly depends on estimating the right mass accretion rate 
which, however, is rather difficult. 

\section{Required magnetic fields in the magnetic jet formation scenario} 
\label{sec_magn}

Let $\Psi$ be a magnetic surface anchored at the inner radius of the accretion 
disk. Then this surface remains always inside the jet up to large distances. 
Using the constants of motion along this surface and the jet parameters, one 
can make detailed estimates following from the scenario.

As magneto-centrifugally driven jets have fast-magnetosonic Mach numbers 
$M_{\rm fm} \sim 3$ \citep[e.g.][]{KLB04}, the total magnetic field in the jet 
should be
\begin{equation} \label{B_jet}
\sqrt{B_{p,{\rm jet}}^2+B_{\varphi,{\rm jet}}^2} = \frac{1}{3}\,\sqrt{4\,\pi\,
\rho_{\rm jet}}\,v_{\rm jet} = 0.34 \, {\rm G} \sim B_{\varphi,{\rm jet}},
\end{equation}
\noindent
as the azimuthal component should be dominant inside the jet. With the 
conservation of current $I = R\,B_{\varphi}$, the azimuthal field near the jet 
source is then
\begin{equation}
B_{\varphi,0} \sim 7.3\,{\rm kG} .
\end{equation}
\noindent
As $B_{\varphi,0} \ll B_{p,0}$ , the total magnetic field should be by far 
larger. An upper limit can be found using again equation (\ref{B_jet}), namely 
$B_{p,{\rm jet}} \ll 0.34$ G, and the conservation of flux $\Psi = B_p\,R^2$, 
leading to magnetic fields in the range of
\begin{equation} \label{B_range}
7.3\,{\rm kG} \ll B_{p,0} \ll 155\,{\rm MG} .
\end{equation}

\section{Observing the magnetic field} \label{sec_obs}

Following \citet{BSK04}, jets are detected in 10 out of almost 200 symbiotic 
stars. Remarkably, this is exactly the fraction of magnetized to un-magnetized 
{\em isolated} white dwarfs with field strengths larger than 30 kG 
\citep{WiF}.

A strong magnetic field, however, has been detected only in one symbiotic 
system \citep[Z Andromedae,][]{SoB99}. They interpreted an observed periodic 
photometric variability as the Keplerian spin period at the magnetospheric 
radius $r_{\rm mag}$, where the magnetic pressure of the WD's field is 
comparable to the ram pressure of the accreted material. This radius can be 
calculated as 
\begin{eqnarray}
r_{\rm mag} &=& \left( \frac{B^4\,R^{12}}{32\,G\,M\,\dot M^2} \right)^{1 / 7}
= 3.6 \times 10^9 \,\left( \frac{B}{{\rm MG}} \right)^{4 / 7}\\
&& \times\,
\left( \frac{R}{7\times 10^8 \,{\rm cm}} \right)^{12 / 7}\,
\left( \frac{M}{M_{\odot}} \right)^{- 1 / 7}\,
\left( \frac{\dot M}{10^{-8}\, M_{\odot} {\rm yr}^{-1}} \right)^{- 2 / 7}\,
{\rm cm} , \nonumber 
\end{eqnarray}
\noindent
which leads to a Keplerian period of
\begin{eqnarray}
P &=& 2\,\pi\,\sqrt{\frac{r_{\rm mag}^3}{G\,M}} = 116 \,
\left( \frac{B}{{\rm MG}} \right)^{6 / 7}\,
\left( \frac{R}{7\times 10^8 \,{\rm cm}} \right)^{18 / 7}\nonumber \\
&& \times\,
\left( \frac{M}{M_{\odot}} \right)^{- 5 / 7}\,
\left( \frac{\dot M}{10^{-8}\, M_{\odot} {\rm yr}^{-1}} \right)^{- 3 / 7}\,
{\rm s} . 
\end{eqnarray}
\noindent
In the case of Z And, the measured period was about 28 min \citep{SoB99}, 
equivalent to a magnetic field of about 22 MG. 

As no oscillations have been detected in MWC 560 so far, the magnetic 
field near the white dwarf should have lower values in the range of eq. 
(\ref{B_range}) -- of the order of 
$B = 100$ kG, with which the period would be 16 s and therefore certainly 
below any observable limit. 

\citet{SoR03}, however, assumed that the magnetic field either inside the jet 
or near the jet source would undergo fast reconnection which heats up the 
plasma to temperatures making the gas visible in X-rays. The maximum 
temperature up to which the gas is heated can be estimated by equipartition 
\citep{TYK03} as
\begin{equation}
T_{\rm max} \sim 10^6 \,\left( \frac{n}{10^{6}\,{\rm cm}^{-3}}\right)^{-1}\,
\left( \frac{B}{100\,{\rm mG}}\right)^{2}\,{\rm K} .
\end{equation}
\noindent
Inside the jet, the maximum temperature, however, would be negligible. Near 
the jet source it would be $\sim 7 \times 10^7$ K with a density of 
$1.4 \times 10^{16}$ cm$^{-3}$ -- which corresponds to an accretion rate of 
$10^{-8}$ $M_{\odot}$ yr$^{-1}$ -- and a magnetic field of 100 kG. The 
emissivity in Bremsstrahlung, created by reconnection, would then be 
$3 \times 10^9$ erg s$^{-1}$ cm$^{-3}$ and, multiplied by the volume of the 
accretion disk affected by the magnetic field
$(2\,\pi\,R_{*})\times(0.1\,R_{*})\times( r_{\rm mag} - R_{*} )$, the total 
luminosity would be $2.4 \times 10^{35}$ erg s$^{-1}$. With again a distance 
of 2.5 kpc, the flux would be $3\times 10^{-8}$ erg s$^{-1}$ cm$^{-2}$. MWC 
560 was not detected by the {\em ROSAT} all-sky survey, although the flux 
limit of {\em ROSAT} is $3 \times 10^{-12}$ erg s$^{-1}$ cm$^{-2}$ 
\citep{CVB02}. 

A common argument in YSOs to explain the non-detection is absorption by a high 
column density. As \citet{SKC01} and \citet{SCS05} have found that the number 
density near the red giant should be of the order of 10$^9$ cm$^{-3}$ and then 
be decreasing following a $1/r^2$-law, the total column density in directions 
with large inclination would be of the order of 10$^{22}$ cm$^{-2}$ which is 
too low to cause the necessary absorption. The column density of the jet in 
MWC 560 has similar values \citep{SKC01} leading to the same conclusion. 
The non-detection therefore sets an upper limit for the magnetic field which is
at the lower end of the range of eq. (\ref{B_range}). 

\section{Discussion}

By analytic estimates for the symbiotic star MWC 560, we have tried to decide 
which of the two main mechanisms -- formation by MHD processes or formation by 
thermal pressure in the boundary layer (BL) -- is more reliable to explain the 
observed parameters of jets in symbiotic stars. This question can not (yet) be 
answered by observations. 

The X-ray observations of MWC 560 giving an upper limit for the magnetic field 
strength show that is still consistent with the minimal magnetic field 
required for a magnetic jet formation process. Further X-ray observations are 
needed to reduce this detection limit and to decide whether the magnetic 
scenario can be excluded or not. Until very high speed photometric 
observations are made, also the first way to deduce the magnetic field 
strength does not reveal any reason against a magnetic jet formation 
process. At the moment, both mechanisms are still possible.

As the BL scenario was introduced only analytically and the derivation of the 
model parameters by observation is highly difficult, however, numerical 
simulations of a compact object with a solid surface accreting matter would be 
interesting. Then the reliability could be qualitatively and quantitatively 
checked in different ranges of parameters, which should be interesting not 
only for symbiotic stars, but also for other kinds of binaries. The 
effectivity of jet launching by thermal pressure inside the BL should be 
investigated and dependencies of jet parameters as outflow rate, outflow 
velocity and jet kinetic luminosity from initial model parameters as e.g. the 
mass accretion rate should be derived. The authors have started first 
simulations whose results will be presented soon.

\begin{acknowledgements}
Parts of this work were supported by the Deutsche Forschungsgemeinschaft (DFG).
We acknowledge the improving comments and suggestions by the referee.
\end{acknowledgements}

\end{document}